# Verification and Invalidation of the Theory of Symplectic Manifold with Contact Degeneracies as Applied to the Classical Field Theory.

*Igor V Sokolov*
University of Michigan, 2455 Hayward Str., Ann Arbor MI48109 USA.

**Abstract.** A theory of Symplectic Manifold with Contact Degeneracies (SMCD) was developed in [1]. The symplectic geometry employs an *anti-symmetric* tensor (*closed differential form*) such as the field tensor used in the classical field theory. The SMCD theory studies *degeneracies* of such form. In [5] the SMCD theory was applied to study a front of an electromagnetic pulsed field propagating into a region with no field. Here, the result of [5] is compared with the problem solution obtained using the Whitham method. It is shown that the SMCD theory prediction differs from the result obtained with the alternative method.

**1. A theory of Symplectic Manifold with Contact Degeneracies** (SMCD) was developed in [1]. Similarly to the Riemann geometry, which constructs a formalism of the general relativity theory [2] based on a *symmetric* metric tensor, the symplectic geometry, in a particular case of 4 Dimensions (4D), employs a 4*4 *anti-symmetric* tensor (*closed differential form*). An example of such form is a field tensor $F^{ij} = (\mathbf{E}, \mathbf{H})$, $\mathbf{E}, \mathbf{H}$, being the electric and magnetic field. This tensor is used to formulate the Classical Field Theory (CFT) [2] in the Minkowski time-space (ct,$\mathbf{x}$), t,$\mathbf{x}$ being the time and 3D Cartesian coordinate vector, c is the speed of light. The condition for the form, $F^{ij}(t, \mathbf{x})$, to be *closed* was shown in [3] to reduce to the first pair of the Maxwell equations [2,4]:

$$\frac{1}{c}\frac{\partial \mathbf{H}}{\partial t} = -[\nabla \times \mathbf{E}], \qquad (\nabla \cdot \mathbf{H}) = 0. \qquad (1)$$

The SMCD theory [1] aims to study *degeneracy points* of a closed differential form, in which its determinant nullifies, under an extra requirement for the degeneracy points to be the *contact* ones. The latter property characterizes a decay rate of the determinant of the differential form while approaching the degeneracy point. The SMCD theory [1] provides results about the contact degeneracies, particularly, the canonical expansion of the differential form near the contact point.

**2. An application of the SMCD theory to the CFT** was considered in [5]. It was noticed therein that the natural example of surface all consisting of degeneracy points is a front of a pulsed ElectroMagnetic (EM) perturbation propagating into a region with no field. In this case the determinant vanishes as long as the field vanishes at a *Null Hyper-Surface* (NHS) separating at each moment of time the spatial region in which the field vanishes identically from the spatial domain in which the non-zero field somehow decays while approaching the NHS. If the field inside the NHS satisfies the equation, derived in [5]:

$$(\mathbf{E}(t,\mathbf{x}_q) \cdot \mathbf{H}(t,\mathbf{x}_q)) = O[\chi^3(t,\mathbf{x}_q)], \quad q \to p \quad , \tag{2}$$

$\mathbf{x}_q$ being the coordinates of an arbitrary point, q, close to the NHS point, p, $\chi(t,\mathbf{x})$ being the distance from an inner point, $\mathbf{x}$, to the NHS at a time t, then this NHS consists of contact points, thus satisfying all assumptions of the SMCD theory. This allowed some predictions in [5] about the EM field structure near the NHS. Particularly, as a consequence from the canonical expansion of the differential form near a contact point the EM field strength was found in [5] to decrease linearly:

$$|\mathbf{E}(t,\mathbf{x}_q)|, |\mathbf{H}(t,\mathbf{x}_q)| = O[\chi(t,\mathbf{x}_q)], \quad q \to p. \tag{3}$$

**3. The SMCD theory as applied to the CFT may be verified** against a benchmark solution for the EM field near the NHS as derived in Section 4 with the Whitham method described in [6]. The assumption that the NHS consists of contact points is discussed in Section 5. It is concluded in Section 6 that the SMCD theory application to the CFT fails.

**4. The Whitham method solves the EM field near NHS** in the way described in [6], section 7.7. The key point is that the fields are not perfectly smooth at the NHS. While outside the NHS the EM field and all its derivatives vanish, at the inner side some field derivatives do not vanish, hence, they are discontinuous across the NHS. Therefore, the solution of the Maxwell equations may be sought in the form as follows: $\mathbf{E} \propto \chi^\alpha$, $\mathbf{H} \propto \chi^\alpha$, $\chi \geq 0$, $\alpha = const > 0$. It is easier to look for an expansion for the vector and scalar potentials, not for the fields:

$$\mathbf{A} = \frac{\chi^{\alpha+1}}{\alpha+1}\mathbf{a}, \quad \Phi = \frac{\chi^{\alpha+1}}{\alpha+1}\varphi, \quad \mathbf{E} = -\frac{1}{c}\frac{\partial \mathbf{A}}{\partial t} - \nabla\Phi, \quad \mathbf{H} = [\nabla \times \mathbf{A}]. \tag{4}$$

A Null Front (NF) consisting of the NHS points at t=const, is a 2D surface in a 3D coordinate space, which bounds the domain of non-zero EM field at the time instant, t. Let $\mathbf{n}$ be the outward directed unit vector normal to the NF. From geometric considerations, gradient of $\chi$ in the inner point, $\mathbf{x}$, equals $\nabla\chi = -\mathbf{n}$, the latter vector is taken at the NF point nearest to $\mathbf{x}$. The time derivative of $\chi$ equals $\frac{\partial \chi}{\partial t} = c$ manifesting the NF propagation with the speed of light. Hence, the fields are as follows: $\mathbf{E} = \chi^\alpha(\mathbf{n}\varphi_0 - \mathbf{a}_0) + O[\chi^{\alpha+1}]$, $\mathbf{H} = -\chi^\alpha[\mathbf{n} \times \mathbf{a}_0] + O[\chi^{\alpha+1}]$. Here, for given t,$\mathbf{x}$, the subscript knot denotes the values of $\mathbf{a},\varphi$ taken at the NF point nearest to $\mathbf{x}$. The second pair of the Maxwell equations for fields in vacuum reads:

$$\frac{1}{c}\frac{\partial \mathbf{E}}{\partial t} = [\nabla \times \mathbf{H}], \quad (\nabla \cdot \mathbf{E}) = 0. \tag{5}$$

The second of Eqs.(5) applied to the above equation for electric field gives[1]: $\varphi_0 = (\mathbf{n} \cdot \mathbf{a}_0)$. Thus, the EM field near the NHS is found:

$$\mathbf{E} = \chi^\alpha [\mathbf{n} \times [\mathbf{n} \times \mathbf{a}_0]] + O[\chi^{\alpha+1}], \quad \mathbf{H} = -\chi^\alpha [\mathbf{n} \times \mathbf{a}_0] + O[\chi^{\alpha+1}] = [\mathbf{n} \times \mathbf{E}] + O[\chi^{\alpha+1}]. \tag{6}$$

Near the NF, in neglecting the high-order terms the relationships between vectors $\mathbf{n}, \mathbf{E}, \mathbf{H}$ are identical to those for the plane EM wave propagating along the direction of $\mathbf{n}$: $(\mathbf{n} \cdot \mathbf{E}) = 0, \mathbf{H} = [\mathbf{n} \times \mathbf{E}]$ (see Eq.(47.4) in [2]). Whitham noticed in [6] that the abovementioned non-smoothness of the field is equivalent to claiming that the high-frequency harmonics mostly contribute to the Fourier-transformed field, the wavelength for such harmonics tending to zero. Therefore, as long as the wavelength of the EM wave is small comparing with the NF inverse curvature, the front curvature is negligible and 3D wave should be locally close to the plane EM wave - exactly what we have in Eqs.(6).

Disappointingly, the paper [5] arrived at quite opposite conclusion, claiming in the final section that the EM field near the NHS is not any analogous to that in the EM wave. Although started from the same Eqs.(4) as used here (with $\alpha = 1$), the mistaken and incomplete algebra in [5] resulted in incorrect Fig.1 and sophisticated formulae from which the similarity to the plane wave field is difficult to figure out.

5. **Verification of the SMCD theory as applied to the CFT** is now straightforward: (1) the value of $\alpha$ (which keeps uncertain so far) should be found at which the NHS consists of the contact points, that is $(\mathbf{E} \cdot \mathbf{H}) = O[\chi^3]$; (2) thus found value should be compared with value of $\alpha = 1$ as predicted by the SMCD theory (see Eq.(3)) and based on the canonical expansion near the contact degeneracy point. Consider the field in vacuum produced by the point source, which is switched on at the time instant, t=0. In this case the equation of the NHS is as follows:

$$\chi = ct - r = 0, \quad r = \|\mathbf{x}\|. \tag{7}$$

At any point outside the NHS Eq.(7), i.e. at r>ct the field vanishes as long as the waves emanated after the field source is switched on have not yet reached this point. For simplicity, let the field source be the combination of electric and magnetic dipoles, $\mathbf{d}(t), \mathbf{m}(t)$, both directed along the constant unit vector, $\mathbf{l}$, and $\mathbf{d}(t), \mathbf{m}(t) = O[t^{2+\alpha}] \cdot \mathbf{l}$. The vector and scalar potentials of the field produced by an electric dipole, $\mathbf{d}(t) = 0$, $t < 0$, $\mathbf{d}(t) = d t^{2+\alpha} \mathbf{l}$, $t \geq 0$, $d = const$, are as follows:

$$\mathbf{A}_d = \frac{\dot{\mathbf{d}}(t-r/c)}{cr}, \quad \Phi_d = \frac{(\mathbf{n}\cdot\mathbf{d}(t-r/c))}{r^2} + \frac{(\mathbf{n}\cdot\dot{\mathbf{d}}(t-r/c))}{cr}, \tag{8}$$

where $\mathbf{n}=\mathbf{x}/r$. The field vanishes at $r \geq ct$. At $r \leq ct$ we have:

$$\mathbf{H}_d = -\frac{\mathbf{n}\times\ddot{\mathbf{d}}(t-r/c)}{c^2 r} - \frac{\mathbf{n}\times\dot{\mathbf{d}}(t-r/c)}{cr^2}, \quad \mathbf{E}_d = \frac{[\mathbf{n}\times[\mathbf{n}\times\ddot{\mathbf{d}}(t-r/c)]]}{c^2 r} +$$
$$+\frac{3\mathbf{n}(\mathbf{n}\cdot\dot{\mathbf{d}}(t-r/c))-\dot{\mathbf{d}}(t-r/c)}{cr^2} + \frac{3\mathbf{n}(\mathbf{n}\cdot\mathbf{d}(t-r/c))-\mathbf{d}(t-r/c)}{r^3}. \tag{9}$$

---

[1] Note, that the first of Eqs.(5) is not used and the second one is only applied at the NHS, so that Eqs.(6) may be also valid with non-zero current and charge densities.

Eq.(9) is the inverse Fourier transformation of Eqs.(72.6-7) in [2]. The field of the magnetic dipole $\mathbf{m}(t) = 0$, $t < 0$, $\mathbf{m}(t) = m t^{2+\alpha} \mathbf{l}$, $t \geq 0$, $m = const$, may be obtained from Eq.(9) by substituting $\mathbf{H}_m = \frac{m}{d}\mathbf{E}_d$, $\mathbf{E}_m = -\frac{m}{d}\mathbf{H}_d$.

For the total field near the NF, i.e. at small positive $(ct - r)$ we have:

$$\mathbf{E}_d + \mathbf{E}_m = \frac{(2+\alpha)(1+\alpha)(t-r/c)^{\alpha}}{c^2 r}\left(d[\mathbf{n}\times[\mathbf{n}\times\mathbf{l}]] + m[\mathbf{n}\times\mathbf{l}]\right) + O[(t-r/c)^{1+\alpha}],$$

$$\mathbf{H}_m + \mathbf{H}_d = \frac{(2+\alpha)(1+\alpha)(t-r/c)^{\alpha}}{c^2 r}\left(m[\mathbf{n}\times[\mathbf{n}\times\mathbf{l}]] - d[\mathbf{n}\times\mathbf{l}]\right) + O[(t-r/c)^{1+\alpha}]. \tag{10}$$

First, we see that at small *(ct-r)* Eqs.(10) agree with Eq.(6) if:

$$\mathbf{a}_0(t,\mathbf{n}) = -\frac{(2+\alpha)(1+\alpha)}{c^{3+\alpha}t}\left(d[\mathbf{n}\times[\mathbf{n}\times\mathbf{l}]] + m[\mathbf{n}\times\mathbf{l}]\right), \tag{11}$$

thus justifying the Whitham method. Now, to verify the SMCD theory, all we still need is to calculate a dot product, $(\mathbf{E}\cdot\mathbf{H})$. While for both dipole fields this product vanishes identically, $(\mathbf{E}_d \cdot \mathbf{H}_d) \equiv 0$, $(\mathbf{E}_m \cdot \mathbf{H}_m) \equiv 0$, for their total we obtain $((\mathbf{E}_d + \mathbf{E}_m)\cdot(\mathbf{H}_m + \mathbf{H}_d)) = \frac{m}{d}(E_d^2 - H_d^2)$ which gives:

$$(\mathbf{E}\cdot\mathbf{H}) = md\frac{(t-r/c)^{2+2\alpha}}{c^2 r^4}\left(2(2+\alpha)(1+\alpha)[\mathbf{n}\times\mathbf{l}]^2 + 4(2+\alpha)^2(\mathbf{n}\cdot\mathbf{l})^2\right) + O[(t-r/c)^{3+2\alpha}]. \tag{12}$$

With the choice of $\alpha = 1/2$ the requirements of the SMCD theory are satisfied as long as Eqs.(1-2) are satisfied for the anti-symmetric field tensor. However, the prediction Eq.(3) based on the canonical expansion near a contact point fails as long as Eq.(3) predicts $\alpha = 1 \neq 1/2$.

6. **Thus, it may be concluded that** the SMCD theory as applied to the CFT fails. With all satisfied requirements for the theory applicability, it cannot properly predict the field behavior neat a contact point of degeneracy. Vise versa, if the field decays linearly near the NHS (for $\alpha = 1$ - see Eq.(10)), as paper [5] states, then the dot product $(\mathbf{E}\cdot\mathbf{H})$ according to Eq.(12) does not obey the requirement as in Eq.(2).

In [5] there are examples of EM field, in which both Eqs.(2,3) are satisfied, however, this seems to be only achievable with impossibly high electric currents near the NF. Specifically, the Maxwell equation with an arbitrary current density, **j**, reads: $[\nabla \times \mathbf{H}] = \frac{4\pi}{c}\mathbf{j} + \frac{1}{c}\frac{\partial \mathbf{E}}{\partial t}$. Together with Eq.(1) this gives: $\frac{1}{c}\frac{\partial}{\partial t}(\mathbf{E}\cdot\mathbf{H}) = (\mathbf{H}\cdot[\nabla\times\mathbf{H}]) - (\mathbf{E}\cdot[\nabla\times\mathbf{E}]) - \frac{4\pi}{c}(\mathbf{j}\cdot\mathbf{H})$. Due to the last term, both Eqs.(2,3) may be satisfied, if $(\mathbf{j}\cdot\mathbf{H}) \neq 0$, $|\mathbf{j}| = O[\chi]$.

Such sort of solutions, however, is not relevant to the CFT, in which the electric current is produced by motion of charged particles, driven by the Lorentz force, $\mathbf{f}_L = e(\mathbf{E} + [\mathbf{v}\times\mathbf{B}]/c)$, e and **v** being the particle charge and velocity (see Eq.(17.5) in [2]). For the solutions in [5], after the time instant, $t_0 = r/c$, when the NF passes the given point, both field and current are assumed to grow linearly in $(t - t_0)$: $|\mathbf{H}|, |\mathbf{E}|, |\mathbf{j}| = O[(t-t_0)]$. However, this would contradict to the particle motional equations. Indeed, to provide the linearly growing contributions to the current density, the charged particles should move with linearly growing speed, hence, with steady state acceleration. The latter could only be caused by a steady state Lorentz force, which is impossible at $|\mathbf{H}|, |\mathbf{E}| = O[(t-t_0)]$. Note, that the realistic motion of charged particles if placed in the field as in Eq.(6) can be easily solved (see problem 47.2 in [2]). The current, $\mathbf{j} = O[\chi^{\alpha+1}]$, they could produce is much weaker than that assumed in [5].

Thus, the SMDC theory fails to describe realistic current-free EM field near the NHS, and, in turn, the fields it predicts seem to involve currents, which cannot be driven by the particle motion in such fields.

**7. I am indebted to** Prof. Mikhail P. Kharlamov, who noticed that the SMCD theory predictions about the degeneracy type are not perfect.

**References.**